\documentclass[preprint]{aastex}
\usepackage{graphicx}

\begin{document}
\title{Konus-Wind observations of the new soft gamma-ray repeater SGR~0501+4516}

\author{R.~L.~Aptekar\altaffilmark{1},
T.~L.~Cline\altaffilmark{2}, D.~D.~Frederiks\altaffilmark{1},
S.~V.~Golenetskii\altaffilmark{1}, E.~P.~Mazets\altaffilmark{1},
V.~D.~Pal'shin\altaffilmark{1}}

\altaffiltext{1}{Ioffe Physico-Technical Institute of the Russian
Academy of Sciences, St. Petersburg, 194021, Russia}
\altaffiltext{2}{Emeritus: NASA's Goddard Space Flight
Center, Greenbelt, MD 20771, USA}


\begin{abstract}
In 2008 August, the new soft gamma-ray repeater SGR~0501+4516 was discovered by
\emph{Swift}. The source was soon confirmed by several groups in space- and
ground-based multi-wavelength observations. In this letter we report the
analysis of five short bursts from the recently discovered SGR, detected with
Konus-Wind gamma-ray burst spectrometer. Properties of the time histories of
the observed events, as well as results of multi-channel spectral analysis,
both in the 20--300~keV energy range, show, that the source exhibits itself as
a typical SGR. The bursts durations are $\lesssim$0.75~s and their spectra
above 20~keV can be fitted by optically-thin thermal bremsstrahlung (OTTB)
model with $kT_{OTTB}$ of 20--40~keV. The spectral evolution is observed, which
resembles the SGR~1627-41 bursts, where a strong hardness-intensity correlation
was noticed in the earlier Konus-Wind observations. The peak energy fluxes of
all five events are comparable to highest those for known SGRs, so a less
distant source is implied, consistent with the determined Galactic anti-center
direction. Supposing the young supernova remnant HB9 (at the distance of
1.5~kpc) as a natal environment of the source, the peak luminosities of the
bursts are estimated to be (2--5)$\times 10^{40}$~erg~s$^{-1}$. The values of
the total energy release, given the same assumptions, amount to (0.6--6)$\times
10^{39}$~erg. These estimations of both parameters are typical for short SGR
bursts.
\end{abstract}

\section{INTRODUCTION}
Soft gamma-ray repeaters (SGRs) were discovered almost 30 years ago
\citep{Mazets1979a, Mazets1979b}. Two main types of bursting emission are
observed. The active stage of the SGRs consists of many short ($<$ 1 second)
bursts of hard X-rays with energy release in the $10^{39}-10^{41}$~erg range.
This activity may last from several days up to over a year, followed by a
quiescent period of up to several years.  The less frequent SGR activity
consists of the rare, giant flare, with an initial pulse having energy release
as high as $2 \times 10^{46}$~erg \citep{Frederiks2007a} followed by a decaying
pulsating tail. At most, only one such giant flare has yet been observed per
SGR source in the $>$~30 years of total monitoring. All SGRs are now known to
be rapidly slowing X-ray pulsars with periods of 2 to 8 seconds and
luminosities of $L\sim10^{35}$~erg~s$^{-1}$. Sharing these characteristics with
anomalous X-ray pulsars (AXPs), the sources are believed to form a rare class
of strongly magnetized isolated neutron stars (NS); see \cite{Mereghetti2008}
for a recent review. Under the assumption of pure magnetic dipole braking, the
values of their period derivative ($\sim 10^{-13}-10^{-10}$~s~s$^{-1}$) imply
an enormous magnetic field strength of $\sim 10^{14}-10^{15}$~G, which is
orders of magnitude larger than those of the radio pulsars. These extreme
magnetic fields are supposed to be an ultimate source of energy for the bright
persistent X-ray emission and the bursting activity of the SGRs and the AXPs.

Only four SGR sources were definitely known, with only one other possibility,
prior to August 22, 2008, when the \emph{Swift} team reported the discovery of
the new soft gamma-ray repeater SGR~0501+4516 \citep{GCN8112,GCN8113}.
\emph{Swift} observed two repeated bursts, localizing their fading X-ray
counterpart to an area of $2.6$ arcsec radius in the direction of the Galactic
anti-center. The SGR nature of this source was confirmed by the discovery of
the $\sim5.76$~s X-ray pulsar \citep{GCN8118} in the Swift-XRT localization
circle. The pulsar spin-down rate $\dot{P} = 1.5(5) \times 10^{-11}$~s~s$^{-1}$
and the inferred dipole magnetic field of $B = 3 \times 10^{14}$~G were soon
estimated \citep{GCN8166}. The refined phase-coherent solution for the combined
$XMM$-$Newton$, $Suzaku$-XIS, and $Swift$-XRT data sets yields the best fit
$P$=5.7620695(1)~s, $\dot{P} = 6.7(1) \times 10^{-12}$~s~s$^{-1}$, and
$\ddot{P} = -1.6(4) \times 10^{-19}$~s~s$^{-2}$ (for the reference epoch MJD
54701.0) \citep{Rea2009}. Multi-wavelength observations reported possible
near-infrared \citep{GCN8126,GCN8159} and optical \citep{GCN8160} counterparts
for the SGR. The continuing SGR activity was observed by several GRB
instruments \citep{ATel1678,GCN8132,GCN8138} to have declined by the end of
August, 2008 \citep{GCN8150}.

During this period the Konus GRB spectrometer onboard the Wind spacecraft
\citep{Aptekar95} detected five SGR~0501+4516 bursts that were sufficiently
intense to enable the triggered observation mode. Observations of the light
curves, spectral analyses and the estimated energies of these events are
presented here, together with comparisons against known SGR behavior previously
observed with the same Konus-Wind instrument.

\section{OBSERVATIONS}

The initial SGR~0501+4516 bursts reported by \emph{Swift} on August 23, 2008
were too weak to enable the Konus-Wind triggered mode, which is optimized for
GRB detection. However, the first triggered Konus event from that source took
place on the same day at 04:47:46.194~s UT. The next four Konus bursts (see
Table~\ref{TableSummary}) were detected within few days, with the last
triggered event on August 26th.

\begin{figure}
\centering
\includegraphics[width=0.7\textwidth]{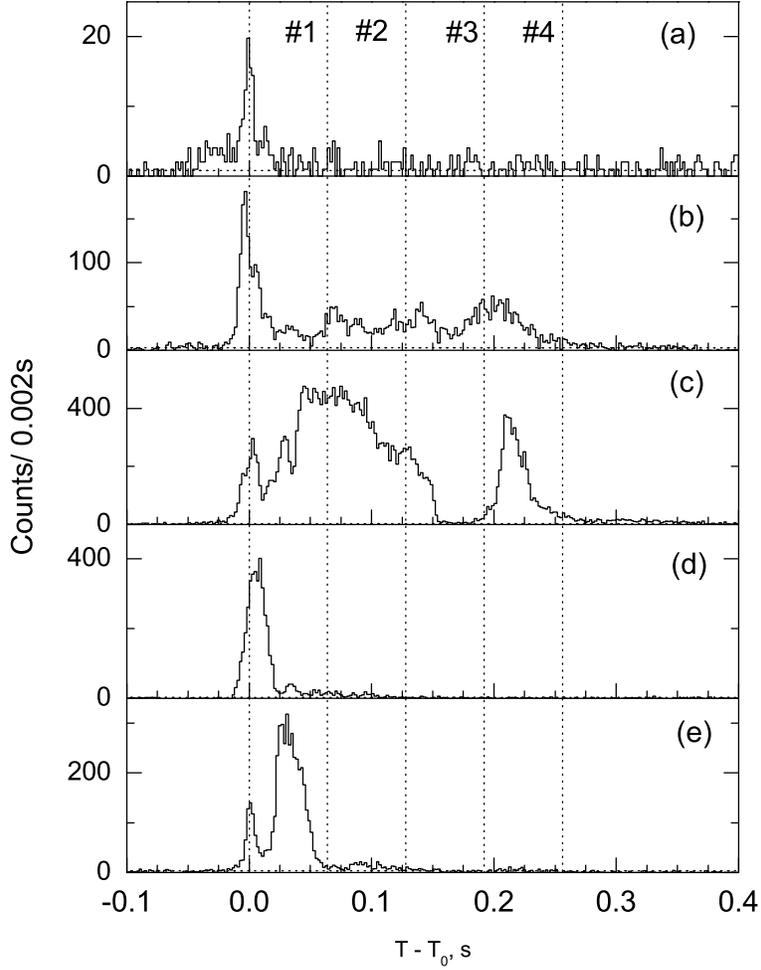}
\caption{The 2~ms light curves of the five SGR~0501+4516 bursts detected by
Konus-Wind. (a)~080823a (in the 70--300~keV range), (b)~080823b, (c)~080824,
(d)~080825, (e)~080826, all in the 20--300~keV range. The trigger times T$_0$
are listed in Table~\ref{TableSummary}. The vertical dashed lines indicate the
four successive 64-ms intervals for which the energy spectra were measured.
\label{allbst}}
\end{figure}

The light curves for these five bursts are shown in Figure~\ref{allbst} with
2~ms resolution. The energy ranges differ in the figures to conform to the
available observations, with the detector's energy thresholds at 20~keV,
70~keV, and 300~keV. Thus, the profile for the 080823a event is for 70 to
300~keV, whereas the other four profiles are for 20 to 300~keV.  No emission
was detected above 300~keV. Dead time corrections of up to 50\% have been made.
It is clear that the bursts are short ($\lesssim$~0.75~s) events and have a
diverse temporal behavior with multiple peaks. The start times, total
durations, and peak times of the bursts are given in Table~\ref{TableSummary}.

For each of these events, up to four energy spectra were measured with 64~ms
accumulation times, as indicated in the event profiles in Figure~\ref{allbst},
illustrated with dashed lines. The raw count-rate spectra were rebinned in
order to have at least 20 counts per energy bin, and fitted using XSPEC,
version 11.3 \citep{Arnaud1996}, using only the 20 to 200 keV fitting interval
since no emission was detected at a higher energy. Results of the spectral
analyses are summarized in Table~\ref{TableSpectra}.

A good fit was obtained for the initial, intense phase to a power-law with an
exponential cutoff (CPL) model, $dN_{ph}/ dE$~$\propto$~$E^{-\alpha} \exp
\left[-(2-\alpha)E/E_p\right]$, where $E_p$ is the peak energy in the $EF(E)$
spectrum. The spectral index values $\alpha$ do not differ far from 1, so the
optically-thin thermal bremsstrahlung (OTTB) model $dN_{ph}/
dE$~$\propto$~$E^{-1} \exp \left[-E/kT\right]$ can be used as well, which is
often considered for SGR spectra above 15~keV. In both cases the values of
$E_p$ (or $kT$) lie in the 20 to 45~keV range. Single and dual black-body
spectral models were also tested, but they are too steep at higher energies to
provide a good agreements to the data above 100 keV.

\begin{figure}
\centering
\includegraphics[width=0.7\textwidth]{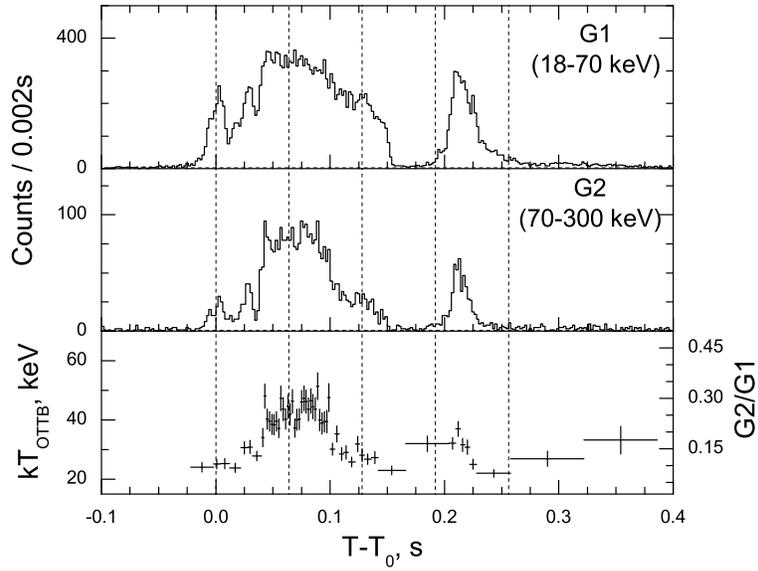}
\caption{The SGR~0501+4516 event detected by Konus-Wind on
August 24, 2008. The two upper panels display the count
profiles in the available energy windows. The lower panel
displays the $kT_{OTTB}$ spectral parameter variation, as
derived from the hardness ratio (right axis). A
hardness-intensity correlation is clearly
apparent.\label{080824}}
\end{figure}

The spectral evolution of the events can be demonstrated, making use of the
hardness ratio of counts in the two available energy windows, G1~(20--70~keV)
and G2~(70--300~keV). The behavior of the G2/G1 hardness ratio and the
$kT_{OTTB}$ spectral parameter, the value of which is estimated using a
detector response matrix, is plotted in Figure~\ref{080824} for the 080824
burst, together with its two channel light curves. A hardness-intensity
correlation is apparent for all observed bursts, and the range of the
$kT_{OTTB}$ variation is in a good agreement with the one derived from the
multichannel spectra.

Using the observed event profiles and spectral analyses,
the total energy fluences and the 2~ms peak fluxes in the
20 to 200 keV range were determined, as summarized in
Table~\ref{TableSummary}. All errors given in
Tables~\ref{TableSummary} and ~\ref{TableSpectra} are at
the 90\% confidence level.

\section{DISCUSSION}
The Konus-Wind omnidirectional detector array constantly observes the whole sky
in a wide energy range from $\sim$20~keV to $\sim$14~MeV. In interplanetary
space far outside the Earth's magnetosphere, Konus has the advantages over
Earth-orbiting GRB monitors of continuous coverage, uninterrupted by Earth
occultation, and a steady background, undistorted by passages through the
Earth's trapped radiation. Therefore, it has had the opportunity to observe
almost all intense bursts from the known Galactic soft gamma-ray repeaters
(SGR~1806-20, SGR~1900+14, and SGR~1627-41) while in their active states
\citep{Aptekar2001}, during the last 14 years of its continuous operation.
Thus, use of the same instrument provides the optimum basis for comparison of
the newly discovered SGR~0501+4516 activity.

The durations, the profiles, and the energy spectra of the SGR~0501+4516 bursts
are all typical for short SGR bursts. The spectral evolution observed in these
events is not in common with SGR~1806-20 or SGR~1900+14 \citep{Aptekar2001},
but resembles the SGR~1627-41 bursts, where a strong hardness-intensity
correlation was noticed in the Konus-Wind observations \citep{Mazets1999}. In
contrast, a hardness-intensity anti-correlation was found in some weak short
bursts as well as throughout the overall burst sample from SGR 1806-20 observed
with INTEGRAL \citep{Gotz2004, Gotz2006}. Among 55 bursts from SGR 1806-20 and
SGR 1900+14 observed by HETE-2 only for 3 bursts was a clear (but moderate)
hardness-intensity anti-correlation reported; for the other 3 bursts, a hint of
a hard component at end of the burst was found \citep{Nakagawa2007}. However,
it should be noted that most of the HETE-2 and INTEGRAL bursts are much weaker
than the Konus-Wind trigger bursts. A similar anti-correlation was found in two
peculiar hard intense bursts from SGR 1900+14 \citep{Woods1999}. The mechanism
responsible for the observed spectral evolution is not yet established.

Since the short bursts from SGR~1627-41 and SGR~0501+4516 are similar to those
from other SGRs in terms of their characteristics such as light curve shapes,
durations, and spectra, they are thus very likely caused by the same physical
process~\footnote{Two such processes in the framework of the magnetar model
have been suggested -- small-scale cracking of a NS
crust~\citep{ThompsonDuncan1995} and reconnection in a NS magnetosphere
\citep{Lyutikov2002, Lyutikov2003}}, and the strong spectral evolution observed
in these two SGRs may be due to some specific characteristics of the sources
(e.g. magnetic field strength and configuration).

\begin{figure}
\centering
\includegraphics[width=0.7\textwidth]{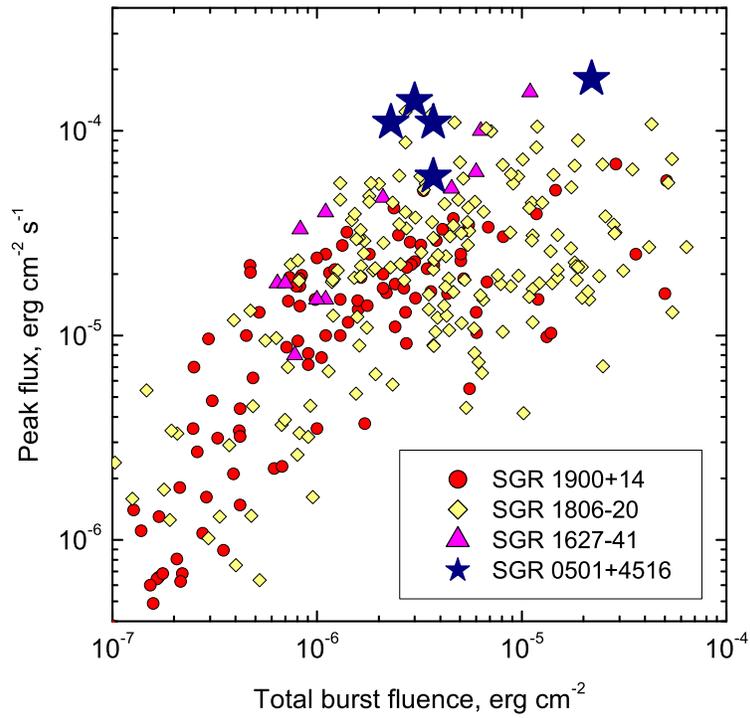}
\caption{The burst energetics, shown with the peak flux vs. the total fluence
for each of the events from 4 SGRs: SGR~0501+4516 (marked by stars), selected
short bursts detected by Konus-Wind from SGR~1806-20 (diamonds), SGR~1900+14
(circles), and SGR~1627-41 (triangles). All the values are in the 20 to 200 keV
range, with the peak fluxes from data on the 2 ms scale. \label{allSGR}}
\end{figure}

The one outstanding feature in the SGR~0501+4516 bursts phenomenology is the
relatively high intensity. The values of the energy fluence for these events
are typically greater than average among the known Galactic SGRs. Even more
distinctive features are the average peak fluxes ($\sim
10^{-4}$~erg~cm$^{-2}$~s$^{-1}$) of the five detected events, at the maximum
range exhibited by all the other SGR bursts observed (Figure~\ref{allSGR}).

The SGR bursts fluences tend to follow a power law distribution
\citep{Gogus2000,Aptekar2001,Gotz2006}. However, the exact shape and parameters
of the observed fluences and peak fluxes distributions are not well determined,
and may be subject to a detection-specific bias. Therefore, model-independent
nonparametric methods were used to test the assumption that SGR~0501+4516
bursts peak fluxes are distributed quantitatively in the same way as those from
the other SGRs detected by Konus-Wind. This hypothesis is rejected for all
three SGR's bursts sets at the 0.05 level by both Kolmogorov-Smirnov and
Mann-Whitney tests. The probability of obtaining a random sample of five bursts
with the measured peak fluxes is very low, it varies from 0.02 for SGR~1627-41
to $\sim10^{-7}$ for SGR~1900+14. In the case of the bursts fluences, the
estimated probabilities are in the 0.36--0.05 range. It should be noted,
however, that these values are affected by the high fluences of a number of
relatively long ``intermediate" bursts (up to tens seconds in duration) which
were not excluded from the reference sets.

Supposing there is no intrinsic difference in the bursts emission mechanism,
the reasonable explanation of the peculiar brightness of the observed
SGR~0501+4516 bursts can be the relatively lesser distance to the source. Such
an interpretation \citep{GCN8138} is supported by the localization of the new
SGR towards the Galactic anti-center. Without well established natal
environments of the known SGRs, their distances are quite uncertain. Various
estimations agree in the conservative lower limits of 5-8~kpc, with more
probable values being closer to 15~kpc for all three known Galactic SGRs
\citep{Hurley1999, Vrba2000, Corbel1999, Corbel2004}. Soon after the
SGR~0501+4516 discovery was found, the proximity of its direction to the young
supernova remnant HB9 was reported \citep{GCN8149}. Assuming a distance of
$\sim$1.5~kpc to HB9 \citep{Leahy1995}, the peak luminosities of the
SGR~0501+4516 bursts observed by Konus-Wind are estimated to be (2--5)$\times
10^{40}$~erg~s$^{-1}$. The values of the total energy release, given the same
assumptions, amount to (0.6--6)$\times 10^{39}$~erg. In conclusion, these
estimations of both parameters are typical for short SGR bursts.

This work was supported by Federal Space Agency of Russia and RFBR grant
09-02-00166a. We gratefully acknowledge an anonymous referee for the detailed
comments and suggestions, which have significantly improved this paper.

\newpage

%

%

%
\begin{deluxetable}{lcccccc}
\tablewidth{0pt}
\tablecaption{Summary of SGR~0501+4516 bursts observed by
Konus-Wind\label{TableSummary}}
\tablehead{\colhead{Burst\tablenotemark{a}} &
\colhead{T$_0$\tablenotemark{b}} &
\colhead{T$_{start}$\tablenotemark{c}} &
\colhead{Duration} &
\colhead{Fluence\tablenotemark{d}} &
\colhead{T$_{peak}$\tablenotemark{e}} &
\colhead{Peak flux\tablenotemark{d}}\\
\colhead{date} &
\colhead{UT} &
\colhead{s} &
\colhead{s} &
\colhead{$10^{-6}$~erg~cm$^{-2}$} &
\colhead{s} &
\colhead{$10^{-4}$~erg~cm$^{-2}$~s$^{-1}$} }
\startdata
080823a & 04:47:46.194 & -0.064 & 0.752 &  2.32$\pm$0.18 & -0.002 & 1.08$\pm$0.44 \\
 & (04:47:48.615) & & &\\
080823b & 11:27:32.652 & -0.068 & 0.446 &  3.72$\pm$0.15 & -0.004 & 0.62$\pm$0.10 \\
 & (11:27:35.042) & & &\\
080824  & 01:17:55.316 & -0.088 & 0.700 & 22.12$\pm$0.68 & 0.074 & 1.76$\pm$0.17 \\
 & (01:17:57.640) & & &\\
080825  & 04:48:27.445 & -0.026 & 0.170 &  2.97$\pm$0.20 & 0.008 & 1.44$\pm$0.17  \\
 & (04:48:29.634) & & &\\
080826  & 03:16:15.017 & -0.166 & 0.514 &  3.74$\pm$0.20 & 0.030 & 1.12$\pm$0.13 \\
 & (03:16:17.094) & & &\\

\enddata
\tablenotetext{a}{Bursts 080823a and 080823b were localized by Swift
\citep{ATel1678}, 080824 and 080825 -- by Fermi \citep{GCN8139}, and 080826 --
by Suzaku \citep{Enoto2009}.}
\tablenotetext{b}{The Konus-Wind trigger time and the corresponding
Earth-crossing time (in the brackets)}
\tablenotetext{c}{The start time of the burst, relative to the trigger time
T$_0$}
\tablenotetext{d}{In the 20--200~keV range, peak fluxes are on the 2~ms scale.}
\tablenotetext{e}{The start time of the 2-ms peak interval, relative to the
trigger time T$_0$}
\end{deluxetable}
\begin{deluxetable}{rllllll}
\tablewidth{0pt} \tablecaption{Summary of spectral fits
\label{TableSpectra}} \tablehead{ \colhead{Burst} & \colhead{Spectra \#\#} &
\colhead{$\alpha$(CPL)} & \colhead{$E_p$(CPL)} & \colhead{$\chi^2$/dof(CPL)} &
\colhead{$kT_{OTTB}$} & \colhead{$\chi^2$/dof(OTTB)}\\
\colhead{} & \colhead{(see Fig.~\ref{allbst})} & \colhead{} & \colhead{(keV)} &
\colhead{} & \colhead{(keV)} & \colhead{}}
\startdata
080823a & 1 & 0.72$_{-0.97}^{+0.88}$ & 26.1$_{-11.0}^{+6.0}$ & 19.3/10 & 23.6$_{-2.5}^{+2.9}$ & 19.6/11\\
& 1-4 & 1.42$_{-0.76}^{+0.56}$ & 18.6$_{-3.5}^{+9.7}$ & 30.6/16 &
24.6$_{-2.4}^{+2.6}$ & 31.5/17\\\hline
080823b & 1 & 1.05$_{-1.04}^{+0.92}$ & 22.7$_{-8.4}^{+7.5}$ &
11.0/10 & 23.2$_{-2.5}^{+2.9}$ & 11.0/11\\
& 1-4 & 0.86$_{-0.36}^{+0.33}$ & 29.8$_{-4.6}^{+3.5}$ & 27.6/20 &
28.1$_{-1.3}^{+1.5}$ & 28.1/21\\\hline
080824 & 1 & 1.17$_{-0.50}^{+0.47}$ & 37.1$_{-8.8}^{+6.1}$ & 21.7/17 & 39.1$_{-3.3}^{+3.7}$ & 22.1/18\\
& 2 & 0.71$_{-0.40}^{+0.36}$ & 45.3$_{-5.1}^{+4.1}$ & 23.8/19 & 41.9$_{-2.7}^{+2.9}$ & 25.6/20\\
& 4 & 1.26$_{-0.56}^{+0.50}$ & 27.3$_{-7.0}^{+7.4}$ & 24.9/15 & 31.1$_{-2.5}^{+2.7}$ & 25.5/16\\
& 1-4 & 0.98$_{-0.22}^{+0.20}$ & 37.9$_{-3.6}^{+3.0}$ & 51.1/24 & 37.7$_{-1.5}^{+1.5}$ & 51.2/25\\
\hline
080825 & 1 & 1.15$_{-0.92}^{+0.84}$ & 28.4$_{-8.3}^{+7.5}$ & 7.5/12 & 29.9$_{-3.4}^{+4.1}$ & 7.6/13\\
& 1-2 & 0.48$_{-0.65}^{+0.59}$ & 31.6$_{-6.4}^{+4.4}$ & 14.3/14 &
26.5$_{-2.0}^{+2.1}$ & 16.3/15\\\hline
080826 & 1 & 1.00$_{-0.54}^{+0.49}$ & 29.5$_{-8.9}^{+5.5}$ & 12.1/15 & 29.6$_{-2.2}^{+2.3}$ & 12.1/16\\
& 1-4 & 0.99$_{-0.49}^{+0.45}$ & 28.1$_{-7.7}^{+5.1}$ & 13.2/19 & 28.0$_{-1.9}^{+2.0}$ & 13.2/20\\
\enddata
\end{deluxetable}
%
\newpage

\newpage

\newpage


\begin{thebibliography}{}

%
\bibitem[Aptekar et al.(1995)]{Aptekar95}Aptekar, R. L., et al. 1995, \ssr, 71, 265
%
\bibitem[Aptekar et al.(2001)]{Aptekar2001}Aptekar, R. L., et al. 2001, \apjs, 137, 222
%
\bibitem[Arnaud (1996)]{Arnaud1996}Arnaud, K. A. 1996, in ASP Conf. Ser. 101,
Astronomical Data Analysis Software and Systems V, ed. G. Jacoby, \& J. Barnes,
(San Francisco: ASP), 17
%
\bibitem[Barthelmy et al.(2008)]{GCN8113}Barthelmy, S. D., et al. 2008, GCN
Circular 8113
%
\bibitem[Corbel et al.(1999)]{Corbel1999}Corbel, S., et al., 1999, ApJ, 526, L29
%
\bibitem[Corbel \& Eikenberry (2004)]{Corbel2004}Corbel, S., and Eikenberry, S. S., 2004,  A\&A, 419, 191
%
\bibitem[Enoto et al.(2009)]{Enoto2009}Enoto, T., et al. 2009, \apj, 693,
L122
%
\bibitem[Fatkhullin et al.(2008)]{GCN8160}Fatkhullin, T., et al. 2008, GCN Circular
8160
%
\bibitem[Fishman et al.(2008)]{GCN8139}Fishman, G. J., et al. 2008, GCN Circular
8139
%
\bibitem[Frederiks et al.(2007a)]{Frederiks2007a}Frederiks, D. D., et al. 2007a, Astronomy Letters, 33,
11, 3
%
\bibitem[Gaensler \& Chatterjee (2008)]{GCN8149}Gaensler, B. M., and Chatterjee, S. 2008, GCN Circular
8149
%
\bibitem[G{\"o}{\u g}{\"u}{\c s} et al.(2000)]{Gogus2000}G{\"o}{\u g}{\"u}{\c s}, E., et al. 2000,
\apj, 532, L121
%
\bibitem[Gogus et al.(2008)]{GCN8118}Gogus, E., et al. 2008, GCN Circular 8118
%
\bibitem[Golenteskii et al.(1984)]{Golen1984}Golentskii, S., Ilinskii, V.,
\& Mazets, E. 1984, \nat, 307, 41
%
\bibitem[Golenetskii et al.(2008)]{GCN8132}Golenetskii, S. V., et al. 2008, GCN Circular
8132
%
\bibitem[G{\"o}tz et al.(2004)]{Gotz2004}G{\"o}tz, D., et al. 2004, A\&A, 417, L45
%
\bibitem[G{\"o}tz et al.(2006)]{Gotz2006}G{\"o}tz, D., et al. 2006, A\&A, 445, 313
%
\bibitem[Holland et al.(2008)]{GCN8112}Holland, S. T., et al. 2008, GCN
Circular 8112
%
\bibitem[Hurley et al.(1999)]{Hurley1999}Hurley, K., et al. 1999, \apj, 510,
L111
%
\bibitem[Kouveliotou et al.(2008)]{GCN8138}Kouveliotou, C., et al. 2008, GCN Circular
8138
%
\bibitem[Krimm et al.(2008)]{GCN8120}Krimm, H. A., et al. 2008, GCN Circular
8120
%
\bibitem[Lyutikov (2002)]{Lyutikov2002}Lyutikov, M. 2002, ApJ, 580, L65
%
\bibitem[Lyutikov (2003)]{Lyutikov2003}Lyutikov, M. 2003, MNRAS, 346, 540
%
\bibitem[Leahy \& Aschenbach(1995)]{Leahy1995}Leahy, D. A., and  Aschenbach, B. 1995, A\&A, 293, 853
%
\bibitem[Mazets et al.(1979a)]{Mazets1979a}Mazets, E. P., et al. 1979a, \nat, 282,
587
%
\bibitem[Mazets et al.(1979b)]{Mazets1979b}Mazets, E. P., et al. 1979b, Sov. Astron. Lett., 5(6), 343
%
\bibitem[Mazets et al.(1999)]{Mazets1999}Mazets, E. P., et al. 1999, \apj, 519, L151
%
\bibitem[Mereghetti (2008)]{Mereghetti2008}Mereghetti, S. 2008, A\&ARv, 15, 4, 225
%
\bibitem[Nakagawa et al.(2007)]{Nakagawa2007}Nakagawa, Y. E., et al. 2007, \pasj, 59, 653
%
\bibitem[Palmer et al.(2008a)]{ATel1678}Palmer, D. 2008, Astronomers's Telegram
1678
%
\bibitem[Palmer et al.(2008b)]{GCN8150}Palmer, D., et al. 2008, GCN Circular
8150
%
\bibitem[Rea et al.(2008a)]{GCN8159}Rea, N., et al. 2008, GCN Circular 8159
%
\bibitem[Rea et al.(2008b)]{GCN8165}Rea, N., et al. 2008, GCN Circular 8165
%
\bibitem[Rea et al.(2009)]{Rea2009}Rea, N., et al. 2009, MNRAS accepted (arxiv:0904.2413)
%
\bibitem[Tanvir \& Varricatt (2008)]{GCN8126}Tanvir, N. R. and Varricatt, W. 2008, GCN Circular 8126
%
\bibitem[Thompson \& Duncan (1995)]{ThompsonDuncan1995}Thompson, C., \& Duncan, R. C. 1995, MNRAS, 275, 255
%
\bibitem[Vrba et al.(2000)]{Vrba2000}Vrba, F.J., et al., 2000, \apj, 533, L17
%
\bibitem[Woods et al.(1999)]{Woods1999}Woods, P., et al. 1999, ApJ, 527, L47
%
\bibitem[Woods et al.(2008)]{GCN8166}Woods, P., et al. 2008, GCN Circular 8166
%

\end{thebibliography}
\end{document}